\documentclass[twocolumn,preprintnumbers,prb,amsmath,amssymb]{revtex4}
\usepackage{color}
\usepackage{graphics}
\usepackage{dcolumn}
\usepackage{bm}

\renewcommand\textfraction 0
\renewcommand\topfraction 1
\renewcommand\bottomfraction 1
\begin{document}
\title{Absence of room temperature ferromagnetism in bulk Mn-doped ZnO} \author{S.
Kolesnik\footnote{Electronic mail: kolesnik@physics.niu.edu} and B. Dabrowski}
\affiliation{Department of Physics, Northern Illinois University, DeKalb, IL 60115}
\date{\today}
\begin{abstract}
Structural and magnetic properties have been studied for polycrystalline Zn$_{1-x}$Mn$_x$O
($x=$0.02, 0.03, 0.05). Low-temperature ($\sim500^{\circ}$C) synthesis leaves unreacted starting
ZnO and manganese oxides. Contrary to a recent report, no bulk ferromagnetism was observed for
single-phase materials synthesized in air at temperatures above $900^{\circ}$C. Single-phase
samples show paramagnetic Curie-Weiss behavior.

\end{abstract}

\maketitle

In order to exploit spins as information carriers in functional spintronics it is necessary to
develop new materials that would exhibit both room temperature ferromagnetism and semiconducting
properties. Recent theoretical predictions of room temperature ferromagnetism in Zn$_{1-x}M_x$O,
where $M=$~Mn ($p$-type),~\cite{Dietl00} or Fe, Co, Ni~\cite{Sato01} motivated the study of this
class of materials. In our recent paper~\cite{Kolesnik04}, we have shown that the ferromagnetic
contribution to the magnetization in polycrystalline Zn$_{1-x}M_x$O can originate from
ferromagnetic impurities. A recent article in \textit{Nature Materials} by Sharma \textit{et
al.}~\cite{Sharma03} reported on observation of ferromagnetism above room temperature in bulk
polycrystalline material and thin films of Zn$_{0.98}$Mn$_{0.02}$O. Sharma \textit{et al.} claim
that such materials are obtained homogeneous and uniform from the low-temperature
(500-$700^{\circ}$C) ceramic processing. Several papers alternatively reported the
presence~\cite{Ueda01,Cho02,Lee02,Norton03a,Prellier03,Norton03b,Ip03,Radovanovic03} or
absence~\cite{Fukumura01,Kim02,Tiwari02,Risbud03,Rode03,Cheng03,Kim04} of high temperature
ferromagnetic ordering in Zn$_{1-x}M_x$O, which is a result of different preparation methods. Here
we show that single-phase Zn$_{1-x}$Mn$_x$O ($x\leqslant 0.05$) can be synthesized in air only at
higher temperatures ($>900^{\circ}$C). Low-temperature synthesis leads to incompletely reacted
mixture of diamagnetic ZnO and magnetic manganese oxides. Single-phase polycrystalline
Zn$_{1-x}$Mn$_x$O is paramagnetic similar to other Mn-containing diluted magnetic semiconductors.

The Zn$_{1-x}$Mn$_x$O ($x\leqslant 0.05$) samples in this study were prepared using a standard
solid-state reaction, similar to that used by Sharma \textit{et al.}~\cite{Sharma03} Mixtures of
ZnO and MnO$_2$, (purity 99.999\%, Johnson Matthey Materials, UK and Alfa Aesar, USA, respectively)
were fired in air at 400$^{\circ}$C for 12 hours, pressed into pellet and annealed at increasing
temperatures ($T_{\rm ann}$) up to 1350$^{\circ}$C. For the high-temperature ($T_{\rm
ann}>900^{\circ}$C) annealing, when Mn starts to substitute for Zn in the material, the samples
were reground and re-pressed before each firing.

Magnetic ac susceptibility and dc magnetization were measured using a Physical Property Measurement
System and a Magnetic Property Measurement System (both Quantum Design) at temperatures up to
400~K. X-ray diffraction (XRD) experiments have been performed using a Rigaku x-ray diffractometer.
Energy dispersive x-ray spectroscopy (EDXS) analysis was performed by a Hitachi S-4700-II scanning
electron microscope. Thermogravimetric analysis was done with a Cahn thermobalance.

\begin{figure}[!]
\resizebox{8.5cm}{!}{\includegraphics{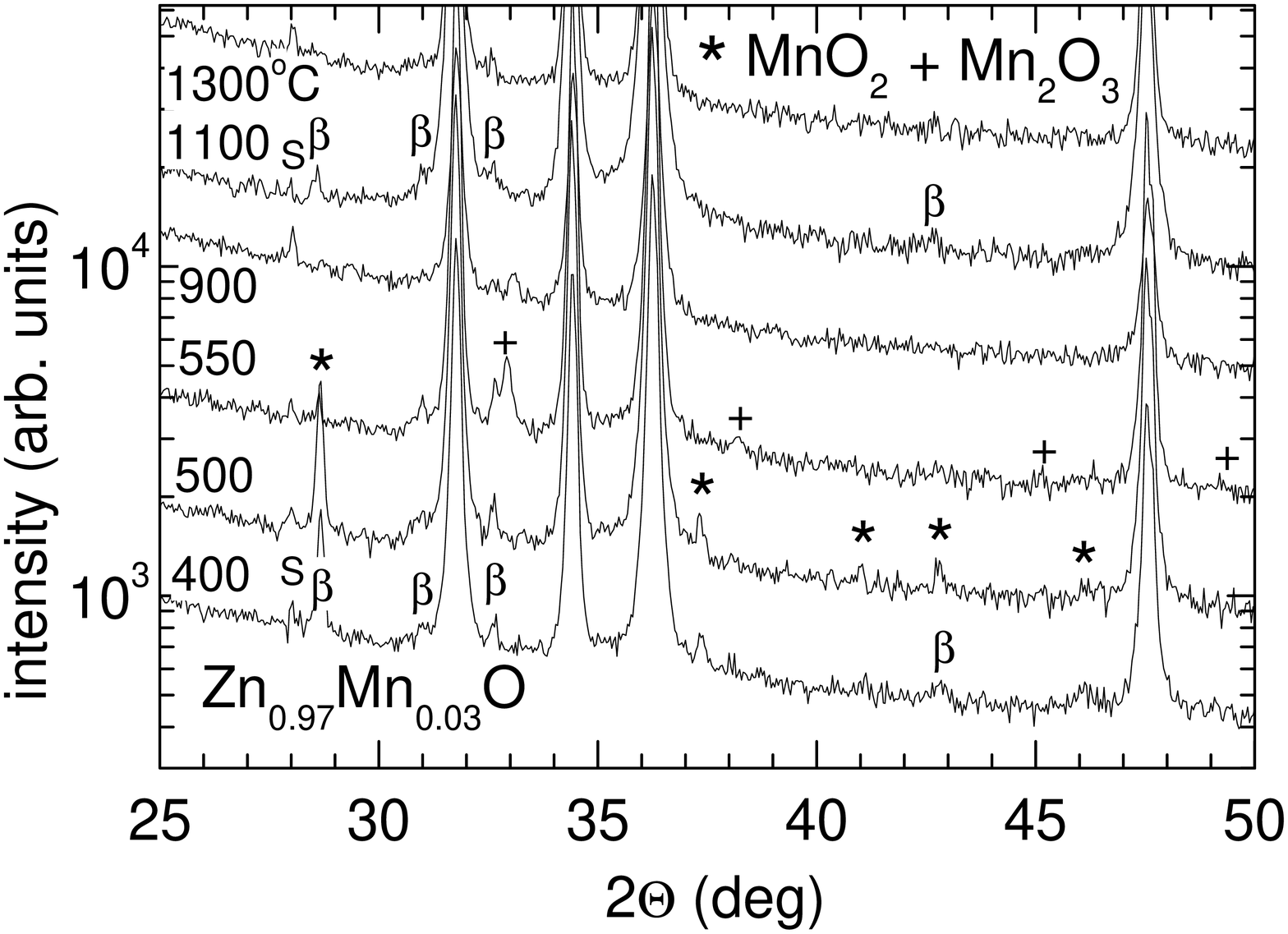}}
\caption{\label{mn003-xray} Semi-logarithmic plots
of x-ray diffraction (XRD) patterns for Zn$_{0.97}$Mn$_{0.03}$O annealed at various temperatures
(listed in the Figure). The spectra are shifted for clarity. Cu${K\beta}$ radiation peaks of ZnO
are marked with $\beta$. Stars and crosses denote impurity peaks of MnO$_2$ and Mn$_2$O$_3$,
respectively. The silica standard peak is marked with 'S'. }
\end{figure}
In Fig.~\ref{mn003-xray}, we show a semi-logarithmic plot of XRD patterns for
Zn$_{0.97}$Mn$_{0.03}$O annealed at various temperatures $T_{\rm ann}$. Essentially, the same
results were obtained for Zn$_{0.98}$Mn$_{0.02}$O. In all the XRD spectra, we observe the main
peaks of the wurtzite structure of ZnO. Each strong peak is accompanied by a smaller peak due to
incompletely filtered Cu${K\beta}$ radiation. Besides these peaks, secondary peaks of manganese
oxides are observed after annealing at low temperatures. For $T_{\rm ann}<500^{\circ}$C, peaks of
MnO$_2$ are observed. For $T_{\rm ann}>500^{\circ}$C, the peaks of Mn$_2$O$_3$ are visible. The
transformation of the secondary phase MnO$_2$ to Mn$_2$O$_3$ in air is consistent with the
structural transition of pure MnO$_2$, observed with XRD and thermogravimetric measurements. At
$T_{\rm ann}=500^{\circ}$C, both manganese oxides are present for Zn$_{1-x}$Mn$_x$O samples
annealed for 12 hours. The presence of manganese oxide peaks is an indication that the
polycrystalline Zn$_{1-x}$Mn$_x$O is not single-phase after low-temperature annealing, contrary of
the conclusion drawn by Sharma \textit{et al.} When a standard ceramic synthesis method is used,
the Zn$_{1-x}$Mn$_x$O compound starts to form a single-phase compound at temperatures higher than
900$^{\circ}$C.

\begin{figure}[!]
\resizebox{8.5cm}{!}{\includegraphics{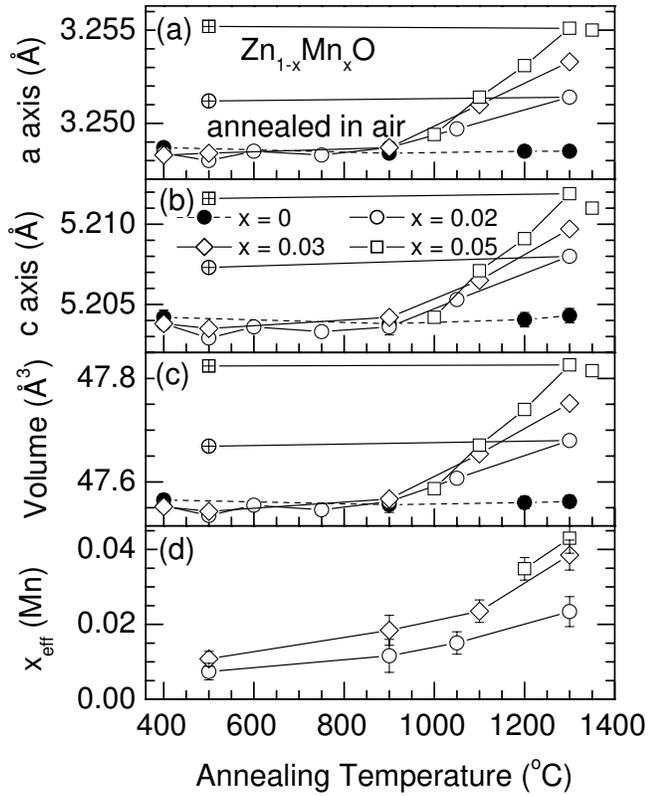}}
\caption{\label{mn00-x-eds} (a-c) Lattice
parameters for Zn$_{1-x}$Mn$_x$O annealed in air at various temperatures, (d) The effective Mn
content determined from the EDXS. Open symbols: substituted samples, filled circles: pure ZnO,
crossed symbols: single-phase samples annealed at 500$^{\circ}$C.}
\end{figure}

The Mn substitution in Zn$_{1-x}$Mn$_x$O can also be easily verified by observing the change of the
lattice parameters as a function of $T_{\rm ann}$.~\cite{Kolesnik04} The lattice parameters of the
wurtzite structure are shown in Figs.~\ref{mn00-x-eds}(a-c), for several nominal Mn contents $x$.
For comparison, the lattice parameters of similarly annealed pure ZnO are also shown.
Fig.~\ref{mn00-x-eds} demonstrates that the lattice parameters of Zn$_{1-x}$Mn$_x$O are fairly
constant and almost identical with those of pure ZnO for $T_{\rm ann}\leqslant 900^{\circ}$C. For
higher $T_{\rm ann}$, the lattice constants gradually increase; this effect is evidence for
substitution of Mn for Zn since Mn$^{2+}$ is larger than Zn$^{2+}$.\cite{Kolesnik04} At $T_{\rm
ann}>900^{\circ}$C, the compositions with $x=0.02-0.05$ become single-phase. Subsequent annealing
of the single-phase samples at 500$^{\circ}$C does not significantly change the lattice parameters,
indicating that the oxidation state of Mn incorporated to Zn$_{1-x}$Mn$_x$O does not change.

\begin{figure}[!]
\resizebox{8.5cm}{!}{\includegraphics{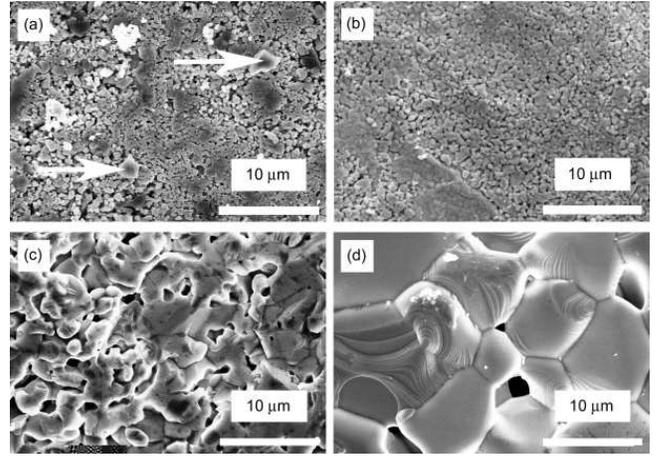}} \caption{\label{mn003eds} Scanning electron
micrographs of Zn$_{0.97}$Mn$_{0.03}$O pellets after annealing in air at (a) $500^{\circ}$C, (b)
$900^{\circ}$C, (c) $1100^{\circ}$C, and (d) $1300^{\circ}$C. Arrows in (a) indicate grains of pure
manganese oxide. }
\end{figure}
In Fig.~\ref{mn003eds}, we show scanning electron micrographs of Zn$_{0.97}$Mn$_{0.03}$O pellets
after annealing at various temperatures. We observe that the size of the grains increases with the
increase of  $T_{\rm ann}$ from less than 1~$\mu$m for $T_{\rm ann}= 500^{\circ}$C up to over
10~$\mu$m for $T_{\rm ann}= 1300^{\circ}$C. EDXS spectra taken on selected areas show the presence
of individual grains of pure manganese oxide for $T_{\rm ann}= 500^{\circ}$C [marked with arrows in
Fig.~\ref{mn003eds}(a)] and pure ZnO. For higher $T_{\rm ann}$, the material is more homogeneous
and more substituted Mn can be detected in the grains. Fig.~\ref{mn00-x-eds}(d) presents the
effective Mn content from the EDXS data. The EDXS data confirm that the Mn substitution gradually
increases with increasing $T_{\rm ann}$ and single-phase Zn$_{1-x}$Mn$_x$O ($x\leqslant 0.05$)
forms at high temperatures ($>900^{\circ}$C).

\begin{figure}[!]
 \resizebox{8.5cm}{!}{\includegraphics{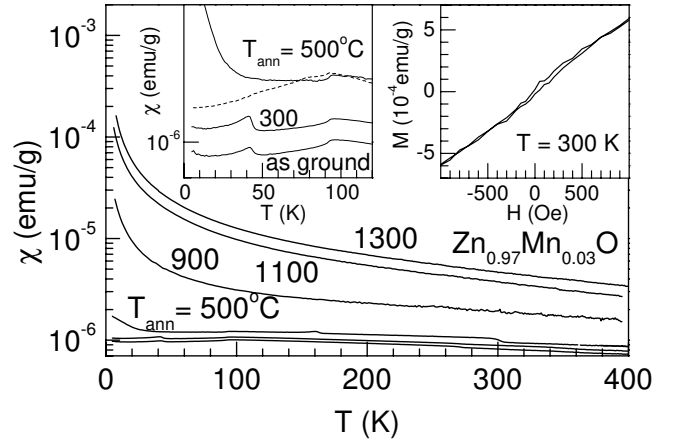}}
\caption{\label{mn003chi} Temperature dependence of the magnetic susceptibility for
Zn$_{0.97}$Mn$_{0.03}$O annealed at various temperatures. Left inset: low-temperature part of the
main panel. Dashed line represents the magnetic susceptibility of MnO$_{2}$ annealed at
500$^{\circ}$C, multiplied by 0.03. Right inset: the field dependence of magnetization of
Zn$_{0.97}$Mn$_{0.03}$O annealed at 500$^{\circ}$C, measured at 300~K.}
\end{figure}
In  Fig.~\ref{mn003chi},  we present temperature dependencies of magnetic susceptibility for the
sample with nominal composition Zn$_{0.97}$Mn$_{0.03}$O annealed at various $T_{\rm ann}$. The
diamagnetic contribution of ZnO, equal to $-0.33\times 10^{-6}$ emu/g, was subtracted from the
measured data. For $T_{\rm ann}<500^{\circ}$C, the susceptibility follows the temperature
dependence expected for manganese oxides that were observed in the XRD spectra. The
antiferromagnetic transition of MnO$_2$ can be seen at $T_N=92$~K and in addition a very small
contribution ($\sim0.01\%$) of ferromagnetic Mn$_{3}$O$_{4}$ with $T_{C}=43$ K is visible. The
Mn$_{3}$O$_{4}$ present in trace amounts in the original MnO$_{2}$ (see: left inset to
Fig~\ref{mn003chi}) shows how apparent vestiges of ferromagnetic impurities are in susceptibility
data. Note that for correct identification of such phases it is critical to perform temperature
dependent measurements. For $T_{\rm ann}= 500^{\circ}$C, the antiferromagnetic transition of
Mn$_{2}$O$_{3}$ is also present at $T_N=76$ K. This behavior is expected for an incompletely
reacted mixture of ZnO and manganese oxides. As a reference, we have also plotted the magnetic
susceptibility for MnO$_{2}$ (dashed line in the left inset to Fig.~\ref{mn003chi}) annealed at
500$^{\circ}$C, multiplied by 0.03. This curve matches well the magnetic susceptibility for
Zn$_{0.97}$Mn$_{0.03}$O for $T_{\rm ann}$=500$^{\circ}$C, except for low temperatures, where a weak
paramagnetic contribution is observed for Zn$_{0.97}$Mn$_{0.03}$O, probably due to a partial
substitution of Mn for Zn in the grain boundaries region. For higher $T_{\rm ann}$, when Mn is
incorporated into the ZnO crystal lattice structure, the material becomes paramagnetic and the
susceptibility increases one hundredfold. This reflects the random distribution of the low
concentration of Mn$^{2+}$ ions on the lattice sites and is characteristic of diluted magnetic
semiconductors. No ferromagnetism at room temperature was detected for any of the studied
Zn$_{1-x}$Mn$_{x}$O samples. Generally, the magnetization shows a linear dependence on the applied
magnetic field. Occasionally, a very small hysteretic contribution to the magnetization can be
observed, but it is always traced to the contamination of the sample holder (see: right inset to
Fig~\ref{mn003chi}).

In summary, we have synthesized polycrystalline Zn$_{1-x}$Mn$_x$O ($x\leqslant 0.05$). This
compound can be formed at temperatures higher than 900$^{\circ}$C using a ceramic route and shows
paramagnetic properties analogous to other diluted magnetic semiconductors. Low-temperature
annealing leaves an incompletely reacted mixture of ZnO and manganese oxides. No bulk
ferromagnetism can be observed for any of the studied samples. At the moment, the origin of the
room temperature ferromagnetic behavior observed by Sharma \textit{et al.}~\cite{Sharma03} is not
clear. Nevertheless, we provide here conclusive evidence that the samples that exhibit
ferromagnetism or antiferromagnetism are not single-phase Zn$_{1-x}$Mn$_x$O compounds.

\acknowledgments

This work was supported by NSF (DMR-0302617), the U.S. Department of Education, and the State of
Illinois under HECA. The EDXS analysis was performed in the Electron Microscopy Center, Argonne
National Laboratory, Argonne, IL.

 \end{document}